\documentclass[preprint]{rsl}

\title{Upper Bounds on Efficiency of Pixel Antennas}
\author{Albert Salmi, Philipp del Hougne, Leo Lahti, and Ville Viikari}

\begin{document}

\maketitle

%
%

\begin{abstract}
This letter presents a method to compute fundamental bounds on the total efficiency of pixel antennas. Using semidefinite relaxation, we obtain an upper bound on the optimization problem of maximizing the antenna's total efficiency subject to binary-valued pixel states. The bounds are demonstrated for antennas with varying numbers of pixels and compared with realizable solutions obtained using the genetic algorithm.
\end{abstract}

\section{Introduction}

Pixel antennas present an optimization-based paradigm for antenna design~\cite{Soltani2018, Jiang2022}. In this approach, the radiating surface is discretized into a set of metallic patches, commonly referred to as pixels. The electrical interconnections between adjacent pixels are binary, corresponding to open or short states, and are determined through numerical optimization to maximize antenna performance with respect to a specified metric. In practical realizations, reconfigurable antenna characteristics, such as operating frequency or radiation pattern, can be achieved by implementing controllable switching elements between the pixels~\cite{Song2014, Rodrigo2016, Tapie2025}.

A pixel antenna with $M$ inter-pixel connections has $2^M$ possible combinations of connection states, corresponding to a wide range of achievable antenna characteristics. Thus, identifying the optimal configuration requires solving a computationally expensive binary optimization problem. Several solution strategies have been proposed, including the genetic algorithm, successive exhaustive Boolean optimization, and a method based on perturbation sensitivity analysis~\cite{Song2014, Shen2017, Jiang2022}. However, obtaining the global optimum -- or even reliably assessing the proximity of a found solution to it -- is rarely feasible within a reasonable computational budget. Thus, it is difficult to determine whether an optimization should be repeated or whether the underlying pixel topology itself should be modified.

\begin{figure}[t]
    \centering
    \includegraphics[width=0.8\linewidth]{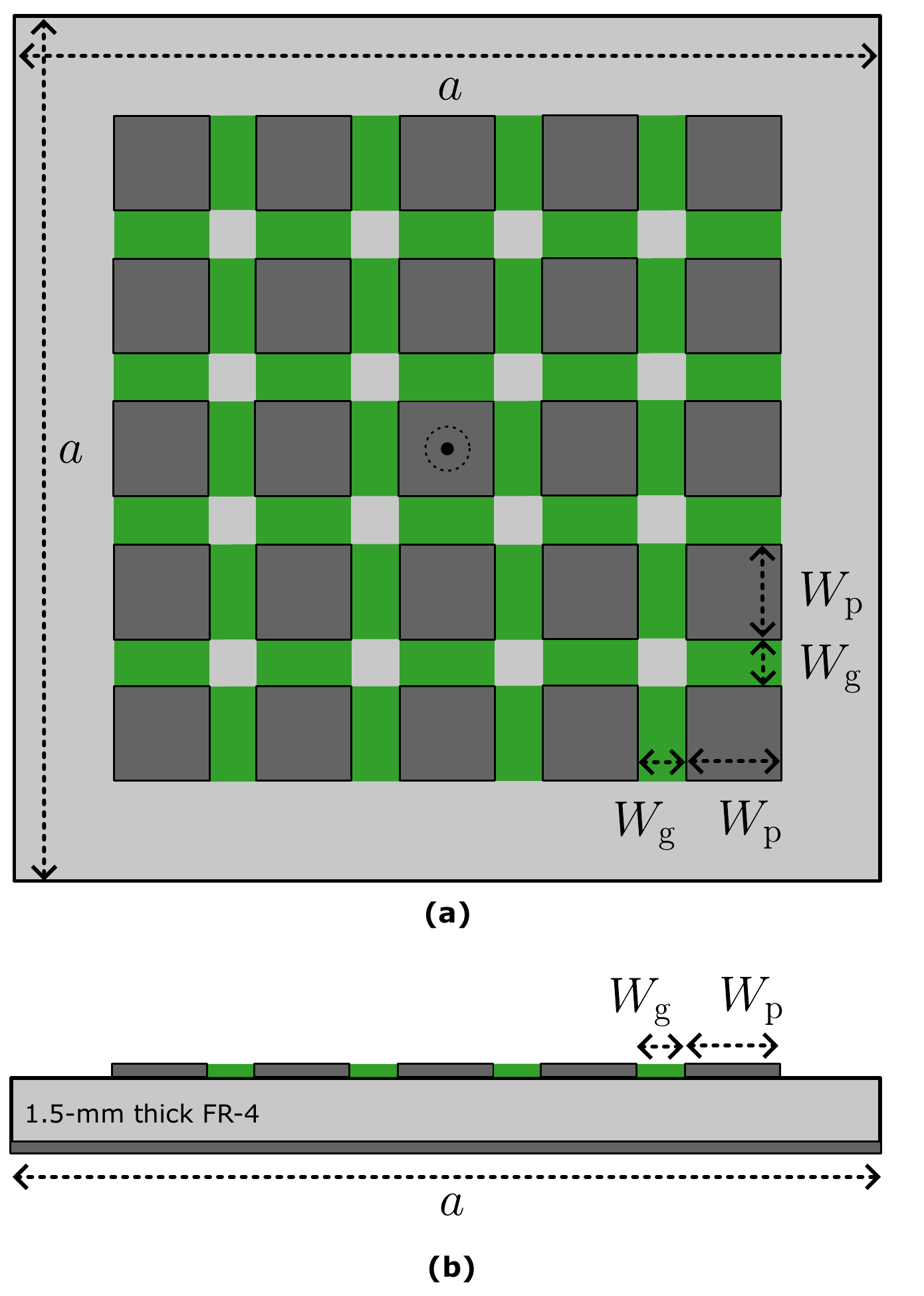}
    \caption{Illustration of a pixel antenna with $P=5$. Light and dark gray indicate the substrate and copper, respectively. Circles mark the coaxial feed, and green rectangles indicate the inter-pixel connections. The side length of the square antenna is $a$, the inter-pixel gap length is $W_{\mathrm{g}}$, and the pixel side length is $W_{\mathrm{p}}$. (a) Top view. (b) Side view.}
    \label{fig:pixelant}
\end{figure}

Fundamental bounds define the performance limits of antennas for a given set of constraints~\cite{Gustafsson2007}. These bounds are unique, computationally efficient to evaluate, and establish absolute limits for the performance metrics under consideration. Analysis of fundamental bounds provides valuable insight into both the suitability of a chosen base antenna topology and the quality of locally optimal solutions obtained through numerical optimization algorithms.

This work derives fundamental upper bounds on the point-frequency total efficiency of pixel antennas using semidefinite relaxation~\cite{Luo2010}. The inherently nonconvex problem of maximizing total efficiency is formulated as a quadratically constrained quadratic program (QCQP), where the optimization variables are the incident waves associated with the pixel interconnections. By relaxing the nonconvex QCQP into a convex semidefinite program, the resulting solution yields an efficiency that exceeds the global optimum of the original problem, without significantly overestimating the achievable performance~\cite{Wang2022}.

The fundamental bounds are analyzed for pixel antennas with the number of pixels varying from $3\times 3$ to $11 \times 11$, and the electrical size of the antenna varying from $0.2\,\lambda$ to $1.6\,\lambda$. The bounds are analyzed alongside realizable solutions computed using the genetic algorithm. 

\section{Total efficiency of pixel antenna} \label{sec:eff}

We study the pixel antenna shown in Fig.~\ref{fig:pixelant}. It contains a copper ground plane, a 1.5-mm-thick FR-4 substrate, and $P \times P$ square copper pixels. The connections, which can be either open or shorted, are placed between the pixels, as indicated in green in Fig.~\ref{fig:pixelant}. The center pixel is fed through a 50-$\Omega$ coaxial connector, with the outer conductor connected to the ground and the inner conductor to the pixel.

The pixel size is $W_{\mathrm{p}} \times W_{\mathrm{p}}$, where
\begin{align}
    W_{\mathrm{p}} = \dfrac{a}{1.3} \dfrac{1}{P + \frac{P-1}{2}}.
\end{align}
The gap between neighboring pixels is $W_{\mathrm{g}} = W_{\mathrm{p}}/2$. The antenna size is $a\times a$, which is $1.3 \times 1.3$ times the area reserved for the pixels and their interconnections.

The antenna is simulated using a full-wave electromagnetic simulation with 50-$\Omega$ discrete ports in the gaps between the pixels. These ports are called \textit{pixel ports}, and referred to with superscript~$\mathrm{P}$. A waveguide port in the coaxial connector is called a \textit{driven port} and referred to with superscript~$\mathrm{D}$. The simulation solves embedded element patterns (EEPs) and scattering parameters (S-parameters) for all ports. 

The antenna's total efficiency is computed based on the EEP of the driven port, which depends on the pixel connection states. The connection between adjacent pixels can be either short or open, resulting in the reflection coefficients $r_m = -1$ or $r_m = 1$, respectively, where the index $m \in \left[1, M\right]$ refers to the connection. The reflection-coefficient-dependent EEP of the driven port is
\begin{align} \label{eq:ehat}
    e_{\alpha}(\theta, \varphi) = e_{\alpha}^{\mathrm{D}}(\theta, \varphi) + (\mathbf{a}^{\mathrm{P}})^{\mathrm{T}} \mathbf{e}_{\alpha}^{\mathrm{P}}(\theta, \varphi),
\end{align}
where $e_{\alpha}^{\mathrm{D}}(\theta, \varphi)$ is the simulated EEP of the driven port, and vector $\mathbf{e}_{\alpha}^{\mathrm{P}}(\theta, \varphi) = \left[ e_{\alpha,1}^{\mathrm{P}}(\theta, \varphi), e_{\alpha,2}^{\mathrm{P}}(\theta, \varphi), \ldots, e_{\alpha,M}^{\mathrm{P}}(\theta, \varphi) \right]^{\mathrm{T}} \in \mathbb{C}^M$ contains the EEPs of the pixel ports, each denoted as $e_{\alpha,m}^{\mathrm{P}}(\theta, \varphi)$~\cite{Salmi2025}. The subscript $\alpha$ defines a polarization, and $(\cdot)^{\mathrm{T}}$ denotes transpose.

The vector $\mathbf{a}^{\mathrm{P}} \in \mathbb{C}^M$ defines the incident waves of the pixel ports, assuming that the incident wave of the feed port has unit amplitude and zero phase, corresponding to the half-watt incident power. The incident waves of the pixel ports depend on the pixel states as~\cite{Jiang2022, Salmi2025}
\begin{align} \label{eq:aP}
    \mathbf{a}^{\mathrm{P}} = (\operatorname{diag}(\mathbf{r})^{-1} - \mathbf{S}^{\mathrm{PP}})^{-1} \mathbf{s}^{\mathrm{PD}},
\end{align}
where $\mathbf{r} \in \{-1, 1\}^M$ is a vector of the pixel port reflection coefficients. The simulated S-parameters $\mathbf{s}^{\mathrm{PD}} \in \mathbb{C}^{M}$ and $\mathbf{S}^{\mathrm{PP}} \in \mathbb{C}^{M \times M}$ define the coupling from the driven port to the pixel ports and between the pixel ports, respectively.

The total efficiency, that is, realized radiation efficiency, is the ratio of radiated power and incident power~\cite{standard}. By fixing the incident power to half a watt, the efficiency can be computed as~\cite{Kildal2015}
\begin{align} \label{eq:eff_orig}
    \eta_{\mathrm{tot}} &= \dfrac{1}{Z_0} \iint_{S} (|e_{\theta}(\theta, \varphi)|^2 + |e_{\varphi}(\theta, \varphi)|^2)\,\mathrm{d}S \nonumber \\
    &\approx \dfrac{4 \pi}{Z_0} \sum_{l=1}^L w_l (|e_{\theta}(\theta_l, \varphi_l)|^2 + |e_{\varphi}(\theta_l, \varphi_l)|^2),
\end{align}
where $Z_0 \approx 377\,\Omega$ is free-space wave impedance and $S$ refers to the sphere. The integration is approximated using a Lebedev quadrature with $L=770$ sampling directions $(\theta_l, \varphi_l)$ with $w_l$ denoting the corresponding quadrature weights~\cite{Lebedev1999}. 

By applying~\eqref{eq:ehat} to~\eqref{eq:eff_orig}, we can write the total efficiency in the following quadratic form:
\begin{align} \label{eq:cost}
    \eta_{\mathrm{tot}} \approx \dfrac{4 \pi}{Z_0}((\mathbf{a}^{\mathrm{P}})^{\mathrm{H}} \mathbf{Q}_0 \mathbf{a}^{\mathrm{P}} + 2 \mathrm{Re}(\mathbf{p}_0^{\mathrm{T}}\mathbf{a}^{\mathrm{P}}) + t_0),
\end{align}
where
\begin{align}
    \mathbf{Q}_0 &= (\mathbf{E}_{\theta}^{\mathrm{P}})^{\mathrm{H}} \operatorname{diag}(\mathbf{w}) \mathbf{E}_{\theta}^{\mathrm{P}} + (\mathbf{E}_{\varphi}^{\mathrm{P}})^{\mathrm{H}} \operatorname{diag}(\mathbf{w}) \mathbf{E}_{\varphi}^{\mathrm{P}}, \\
    \mathbf{p}_0^{\mathrm{T}} &= (\mathbf{e}_{\theta}^{\mathrm{D}})^{\mathrm{H}} \operatorname{diag}(\mathbf{w}) \mathbf{E}_{\theta}^{\mathrm{P}} + (\mathbf{e}_{\varphi}^{\mathrm{D}})^{\mathrm{H}} \operatorname{diag}(\mathbf{w}) \mathbf{E}_{\varphi}^{\mathrm{P}}, \\
    t_0 &= (\mathbf{e}_{\theta}^{\mathrm{D}})^{\mathrm{H}} \operatorname{diag}(\mathbf{w}) \mathbf{e}_{\theta}^{\mathrm{D}} + (\mathbf{e}_{\varphi}^{\mathrm{D}})^{\mathrm{H}} \operatorname{diag}(\mathbf{w}) \mathbf{e}_{\varphi}^{\mathrm{D}},
\end{align}
and $(\cdot)^{\mathrm{H}}$ denotes the conjugate transpose. Matrices $\mathbf{E}^{\mathrm{P}}_{\alpha} = \left[\mathbf{e}_{\alpha}^{\mathrm{P}}(\theta_1, \varphi_1), \mathbf{e}_{\alpha}^{\mathrm{P}}(\theta_2, \varphi_2), \ldots, \mathbf{e}_{\alpha}^{\mathrm{P}}(\theta_L, \varphi_L) \right]^{\mathrm{T}} \in \mathbb{C}^{L \times M}$ with $\alpha \in \{\theta, \varphi\}$ contain the EEPs of pixel ports at all sampling directions $(\theta_l, \varphi_l)$. Similarly, vectors $\mathbf{e}^{\mathrm{D}}_{\alpha} = \left[ e_{\alpha}^{\mathrm{D}}(\theta_1, \varphi_1), e_{\alpha}^{\mathrm{D}}(\theta_2, \varphi_2), \ldots, e_{\alpha}^{\mathrm{D}}(\theta_L, \varphi_L) \right]^{\mathrm{T}} \in \mathbb{C}^L$ represent the EEP of the driven port. The quadrature weights are in vector $\mathbf{w} \in \mathbb{C}^L$. Note that the approximation in~\eqref{eq:cost} comes only from the numerical integration scheme.

\section{Fundamental bound on efficiency} \label{sec:bound}

We consider the optimization problem of finding the reflection-coefficient vector $\mathbf{r} \in \{-1, 1\}^M$ that maximizes~\eqref{eq:cost}. This section formulates the problem as a QCQP in terms of $\mathbf{a}^{\mathrm{P}}$ by expressing the binary constraints as quadratic equality constraints on $\mathbf{a}^{\mathrm{P}}$. The nonconvex QCQP is then relaxed to a convex form via semidefinite relaxation. The solution to the semidefinite program yields the upper bound on the efficiency because the solution to the convex maximization problem has a higher or equal objective function value than the original problem~\cite{Wang2022}.

Next, we express the feasible domain $\mathbf{r} \in \{-1, 1\}^M$ as a set of quadratic equality constraints on $\mathbf{a}^{\mathrm{P}}$. From~\eqref{eq:aP}, the relation between the incident wave at the $m$-th pixel port, $a^{\mathrm{P}}_m$, and the corresponding reflection coefficient~$r_m$ is given by
\begin{align}
    r_m(\mathbf{b}_m^{\mathrm{T}} \mathbf{a}^{\mathrm{P}} + c_m) = a^{\mathrm{P}}_m,
\end{align}
where $\mathbf{b}_m^{\mathrm{T}} = \left[ \mathbf{S}^{\mathrm{PP}} \right]_{m,:}$ is the $m$-th row of $\mathbf{S}^{\mathrm{PP}}$, and $c_m = \left[ \mathbf{s}^{\mathrm{PD}} \right]_m$. Since $r_m = \pm 1$, this relation implies
\begin{align} \label{eq:const_pm}
    \mathbf{b}_m^{\mathrm{T}} \mathbf{a}^{\mathrm{P}} + c_m = \pm a^{\mathrm{P}}_m, \quad \forall m=1,\ldots,M.
\end{align}
Equation~\eqref{eq:const_pm} can equivalently be expressed as
\begin{align} \label{eq:const_prod}
    (\mathbf{b}_m^{\mathrm{T}} \mathbf{a}^{\mathrm{P}} + c_m - a^{\mathrm{P}}_m)^* (\mathbf{b}_m^{\mathrm{T}} \mathbf{a}^{\mathrm{P}} + c_m + a^{\mathrm{P}}_m) = 0, \quad \forall m,
\end{align}
which yields the quadratic form
\begin{align}\label{eq:const_quad}
    (\mathbf{a}^{\mathrm{P}})^{\mathrm{H}} \mathbf{Q}_m \mathbf{a}^{\mathrm{P}} + (\mathbf{a}^{\mathrm{P}})^{\mathrm{H}} \mathbf{p}_{1m} + \mathbf{p}_{2m}^T\mathbf{a}^{\mathrm{P}} + t_m = 0, \quad \forall m,
\end{align}
where
\begin{align}
    \mathbf{Q}_m &= (\mathbf{b}_m^* - \mathbf{u}_m)(\mathbf{b}_m^{\mathrm{T}} + \mathbf{u}_m^{\mathrm{T}}), \\
    \mathbf{p}_{1m} &= (\mathbf{b}_m^* - \mathbf{u}_m) c_m, \\
    \mathbf{p}_{2m}^{\mathrm{T}} &= c_m^*(\mathbf{b}_m^{\mathrm{T}} + \mathbf{u}_m^{\mathrm{T}}), \\
    t_m &= c_m^* c_m.
\end{align}
Here, $\mathbf{u}_m = \left[ \mathbf{I}_M \right]_m$ denotes the $m$-th column of the $M\times M$ identity matrix, and $(\cdot)^*$ denotes the complex conjugation.

The resulting problem is to find~$\mathbf{a}^{\mathrm{P}}$ that maximizes~\eqref{eq:cost} subject to the quadratic equality constraints~\eqref{eq:const_quad}. The problem is non-convex due to a concave objective and non-affine equality constraints. To obtain a tractable reformulation, we introduce the lifted matrix variable~$\mathbf{A}^{\mathrm{P}} = \mathbf{a}^{\mathrm{P}} (\mathbf{a}^{\mathrm{P}})^{\mathrm{H}}$ and express the quadratic terms as
\begin{align}
    (\mathbf{a}^{\mathrm{P}})^{\mathrm{H}} \mathbf{Q} \mathbf{a}^{\mathrm{P}} = \operatorname{trace}(\mathbf{Q} \mathbf{a}^{\mathrm{P}} (\mathbf{a}^{\mathrm{P}})^{\mathrm{H}}) = \operatorname{trace}(\mathbf{Q} \mathbf{A}^{\mathrm{P}}).
\end{align}
With this lifting, the non-convexity is confined to the rank-one constraint~$\mathbf{A}^{\mathrm{P}} = \mathbf{a}^{\mathrm{P}} (\mathbf{a}^{\mathrm{P}})^{\mathrm{H}}$. Replacing this constraint by~$\mathbf{A}^{\mathrm{P}} \succeq \mathbf{a}^{\mathrm{P}} (\mathbf{a}^{\mathrm{P}})^{\mathrm{H}}$ yields the following convex semidefinite program~\cite{Luo2010}:
\begin{align}
\label{eq:sdr}
&\underset{\mathbf{A}^{\mathrm{P}},\, \mathbf{a}^{\mathrm{P}}}{\text{maximize}}
\quad \operatorname{trace}(\mathbf{Q}_0 \mathbf{A}^{\mathrm{P}})
+ 2 \mathrm{Re}(\mathbf{p}_0^{\mathrm{T}}\mathbf{a}^{\mathrm{P}})
+ t_0 \nonumber\\
&\text{subject to}  \\
&\operatorname{trace}(\mathbf{Q}_m \mathbf{A}^{\mathrm{P}})
+ (\mathbf{a}^{\mathrm{P}})^{\mathrm{H}} \mathbf{p}_{1m}
+ \mathbf{p}_{2m}^T \mathbf{a}^{\mathrm{P}}
+ t_m = 0, \quad \forall m, \nonumber\\
&\begin{bmatrix}
\mathbf{A}^{\mathrm{P}} & \mathbf{a}^{\mathrm{P}} \\
(\mathbf{a}^{\mathrm{P}})^{\mathrm{H}} & 1
\end{bmatrix}
\succeq 0. \nonumber
\end{align}

Solution to the relaxed problem yields $(\mathbf{A}^{\mathrm{P}})^{\star}$ and $(\mathbf{a}^{\mathrm{P}})^{\star}$. However, due to the relaxation, the rank-one constraint is not necessarily satisfied. Therefore, feasible binary-valued reflection coefficients cannot be directly obtained from the relaxed solution. Nevertheless, the upper bound on the total efficiency can be solved as
\begin{align}
    \tilde{\eta}_{\mathrm{tot}} = \dfrac{4 \pi}{Z_0}(\operatorname{trace}(\mathbf{Q}_0 (\mathbf{A}^{\mathrm{P}})^{\star}) + 2 \mathrm{Re}(\mathbf{p}_0^{\mathrm{T}}(\mathbf{a}^{\mathrm{P}})^{\star}) + t_0).
\end{align}

\section{Demonstration} \label{sec:demonstration}

\begin{figure}[t]
    \centering
    \includegraphics[width=0.95\linewidth]{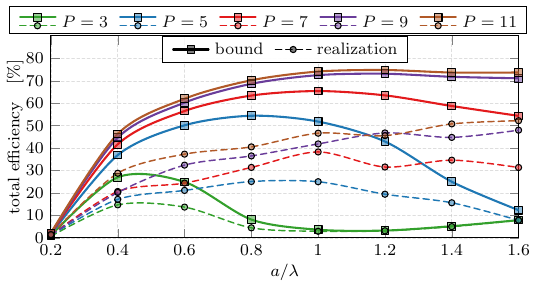}
    \caption{Fundamental bounds and realizable total efficiencies of the studied pixel antenna as a function of antenna size relative to the wavelength.}
    \label{fig:eff_f}
\end{figure}

\begin{figure}[h]
    \centering
    \includegraphics[width=0.95\linewidth]{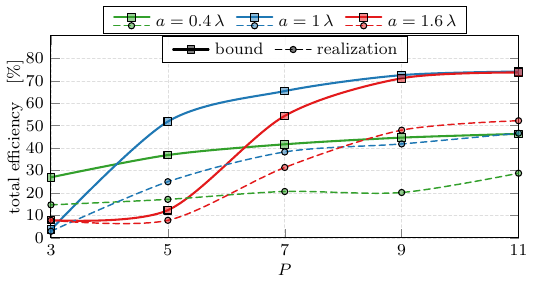}
    \caption{Fundamental bounds and realizable total efficiencies as a function of the number of pixels in a row.}
    \label{fig:eff_P}
\end{figure}

We compute the fundamental bounds on the total efficiency for the pixel antenna shown in Fig.~\ref{fig:pixelant}. The number of pixels per row is swept from $P=3$ to $P=11$, with only odd values considered to retain a central pixel for the coaxial feed. The antenna's electrical size is swept from $a=0.2\,\lambda$ to $a=1.6\,\lambda$. Efficiency is optimized separately for each value of $a$ and $P$. The EEPs and S-parameters of the driven and pixel ports are obtained via frequency-domain simulations that account for the losses in the copper and FR-4.

In addition to computing the fundamental bounds, realizable solutions are also computed. For $P=3$, the realization is obtained by brute-force search for the global optimum. For antennas with more pixels, the realization is computed using the genetic algorithm (GA). In that case, the realization does not necessarily yield the best performance. 

Fig.~\ref{fig:eff_f} shows the total efficiency as a function of the electrical size of the antenna. Solid lines correspond to the fundamental bounds, and dashed lines show realizable efficiencies. Different pixel counts are indicated by colors. Fig.~\ref{fig:eff_P} presents the bound and realizable efficiencies as a function of~$P$ with three electrical sizes, indicated by colors.

Electrically small pixel antennas are capable of radiating only a minor part of the incident power, as indicated in Fig.~\ref{fig:eff_f} and Fig.~\ref{fig:eff_P}. The upper bound on efficiency increases with antenna size until the pixel size becomes a significant fraction of the wavelength. The efficiency bound peaks when $W_{\mathrm{p}} \approx 0.1\,\lambda$, which is consistent with~\cite{Rodrigo2016}. The potential efficiency of sparsely discretized pixel antennas ($P\leq 7$) decreases when the pixel size grows above~$\lambda/10$.

If the discretization is dense enough ($P\geq9$), the upper bound on efficiency peaks at~$a=1.2\,\lambda$. Efficiency does not decrease significantly, although the pixels become electrically larger. A total efficiency greater than 75\% cannot be achieved with this antenna type at any of the studied frequencies or pixel counts. 

Considering a practical antenna design, the bounds indicate that $9\times 9$ pixels are enough, and antenna size of $a = 1.2\,\lambda$ is sufficient to achieve maximum efficiency. This is not obvious from the realizable results, which highlights the importance of examining the fundamental bounds.

These findings are partly generalizable. For instance, the results verify the well-known rule of thumb that the pixel size should be less than~$\lambda/10$, and that the efficiency of small antennas increases with the antenna size. Nevertheless, the fundamental bounds also reveal antenna-specific details. For instance, we learned how many pixels and how large an antenna is required to achieve the maximum efficiency using this antenna base topology. With another substrate, feeding mechanism, or pixel topology, these results could be different. 

\begin{figure}
    \centering
    \includegraphics[width=1\linewidth]{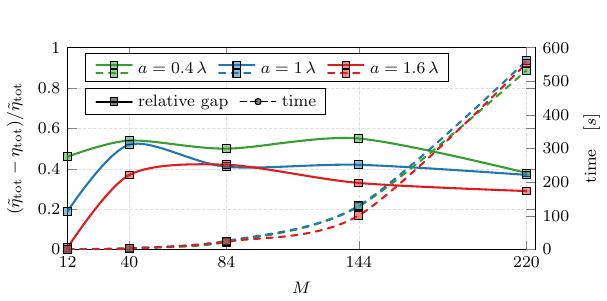}
    \caption{Relative gap between the bound and realizable efficiencies (solid lines) and the time required to compute the fundamental bound (dashed lines) as a function of the number of optimization variables.}
    \label{fig:t_P}
\end{figure}

The solid lines of Fig.~\ref{fig:t_P} present the relative gap between bound efficiency~$\tilde{\eta}_{\mathrm{tot}}$ and GA-optimized efficiency~${\eta}_{\mathrm{tot}}$ as a function of the number of inter-pixel connections~$M$. This metric expresses how much the fundamental bound overestimates the antenna's achievable performance. Note that the suboptimality of the GA solution also affects this metric, resulting in fluctuation in the curves. Thus, the metric does not strictly define the tightness but estimates it. 

Often, the tightness of the semidefinite relaxation, which estimates the gap in Fig.~\ref{fig:t_P}, can be constant or depend on the number of constraint functions, which corresponds to the number of inter-pixel connections in this case~\cite{Luo2010}. The results presented suggest that the tightness of our relaxation is independent of the number of pixels. In addition, the antenna's electrical size does not significantly affect the gap. Thus, the presented fundamental bound is relatively tight also for densely discretized antennas. However, further studies are required to verify that the tighness result also applies to pixel antennas with different base topologies.

Fig.~\ref{fig:t_P} also presents the time required to compute the bounds as a function of the number of connections with dashed lines. The computational complexity of a semidefinite program follows a polynomial in $M$ rather than growing exponentially with $M$, which can be considered as computationally inexpensive~\cite{Luo2010}. Although the complexity cannot be reliably estimated based on Fig.~\ref{fig:t_P}, the figure reveals that the complexity does not depend on the electrical size of the antenna. 

In addition to total efficiency, fundamental bounds of pixel antennas can be computed on several other quantities using the same method. For example, the realized gain and $s_{11}$ at the feed port can be written in the form of~\eqref{eq:cost}, yielding optimization problems similar to that in this work. Extending the method to multi-feed pixel antennas is also possible using a vectorized QCQP formulation similar to that in~\cite{Salmi2025}.

Although this work focuses on a pixel antenna with a regular rectangular layout, the proposed method is directly applicable to a broad class of binary-loaded antenna and scatterer design problems. One such application is the design of chaotic-cavity–based dynamic metasurface antennas~\cite{Tapie2025}.

\section{Conclusions}
This letter proposed a method for computing upper bounds on the total efficiency of pixel antennas via semidefinite relaxation. Analysis of the fundamental bounds revealed valuable information about the potential performance of pixel antennas with different numbers of pixels and varying electrical sizes.

\section{Acknowledgements}

The authors acknowledge the use of MIDAS infrastructure of Aalto 
School of Electrical Engineering. This work was supported by Business Finland through the Multiracs project.


\bibliographystyle{IEEEtran}
\bibliography{Arxiv_references}

%
%
\newpage
\noindent\small
Albert Salmi is with Department of Applied Physics, Aalto University, 02150, Espoo, Finland; e-mail: albert.salmi@aalto.fi.\\[1ex]
Philipp del Hougne is with Department of Electronics and Nanoengineering, Aalto University, 02150, Espoo, Finland, and Univ Rennes, CNRS, IETR -- UMR 6164, F-35000, Rennes, France\\[1ex]
Leo Lahti and Ville Viikari are with Department of Electronics and Nanoengineering, Aalto University, 02150, Espoo, Finland.

\end{document}